\documentclass[twoside]{article}
\usepackage{qic,amsmath,graphicx}
\usepackage[caption=false]{subfig}

\newcommand{\braket}[2]{{\left\langle #1 \middle| #2 \right\rangle}}

\newcommand{\ket}[1]{{\left| #1 \right\rangle}}
\newcommand{\ketbra}[2]{{\left| #1 \middle\rangle \middle \langle #2 \right|}}

\begin{document}

\setlength{\textheight}{8.0truein}    

\runninghead{Correcting for Potential Barriers in Quantum Walk Search}
            {A.~Ambainis, T.~G.~Wong}

\normalsize\textlineskip
\thispagestyle{empty}
\setcounter{page}{1}

\copyrightheading{0}{0}{0000}{000--000}

\vspace*{0.88truein}

\alphfootnote

\fpage{1}

\centerline{\bf CORRECTING FOR POTENTIAL BARRIERS}
\vspace*{0.035truein}
\centerline{\bf IN QUANTUM WALK SEARCH}
\vspace*{0.37truein}
\centerline{\footnotesize ANDRIS AMBAINIS,\enskip THOMAS G.~WONG}
\vspace*{0.015truein}
\centerline{\footnotesize\it Faculty of Computing, University of Latvia, Rai\c{n}a bulv.~19}
\baselineskip=10pt
\centerline{\footnotesize\it R\=\i ga, LV-1586, Latvia}
\vspace*{0.225truein}
\publisher{(received date)}{(revised date)}

\vspace*{0.21truein}

\abstracts{A randomly walking quantum particle searches in Grover's $\Theta(\sqrt{N})$ iterations for a marked vertex on the complete graph of $N$ vertices by repeatedly querying an oracle that flips the amplitude at the marked vertex, scattering by a ``coin'' flip, and hopping. Physically, however, potential energy barriers can hinder the hop and cause the search to fail, even when the amplitude of not hopping decreases with $N$. We correct for these errors by interpreting the quantum walk search as an amplitude amplification algorithm and modifying the phases applied by the coin flip and oracle such that the amplification recovers the $\Theta(\sqrt{N})$ runtime.}{}{}

\vspace*{10pt}

\keywords{Grover's algorithm, quantum walk, amplitude amplification, quantum search, quantum tunneling}
\vspace*{3pt}
\communicate{to be filled by the Editorial}

\vspace*{1pt}\textlineskip


\section{Introduction}

Most quantum algorithms developed to date are based on four general techniques: quantum Fourier transforms (\textit{e.g.}, the Deutsch-Jozsa algorithm \cite{DJ1992} and Shor's algorithm \cite{Shor1997}), amplitude amplification (\textit{e.g.}, Grover's algorithm \cite{Grover1996} and quantum counting \cite{BHT1998}), quantum simulation (\textit{e.g.}, approximating the Jones polynomial \cite{AJL2006} and solving linear equations \cite{HHL2009}), and quantum walks (\textit{e.g.}, element distinctness \cite{Ambainis2004} and NAND evaluation \cite{FGG2008}). Quantum walks are the quantum analogues of classical random walks or Markov chains \cite{ADZ1993,Ambainis2003,Kempe2003}, and their crucial role in many quantum algorithms has spurred tremendous experimental effort to realize them \cite{BMKSW1999,RLBL2005,PLPSMS2008,KFCSAMW2009,SMSGEHS2009}. These physical implementations are not ideal, however, and so substantial theoretical work has investigated the effects of noise and errors in quantum walk algorithms \cite{SBW2003,GKJ2007,Kendon2007,Wong2015c}.

\begin{figure}
\begin{center}
	\subfloat[]{
		\includegraphics{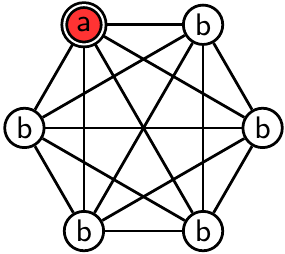}
		\label{fig:complete}
	} \quad
	\subfloat[]{
		\includegraphics{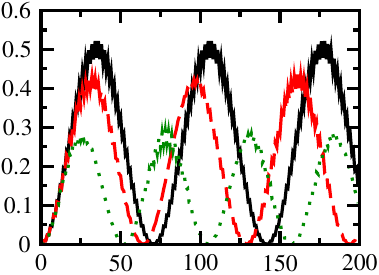}
		\label{fig:prob_time_N1024} 
	}
	\vspace*{13pt}
	\fcaption{(a) The complete graph with $N = 6$ vertices. A vertex is marked, as indicated by a double circle. Identically evolving vertices are identically colored and labeled. (b) Success probability as a function of the number of applications of $U$ for search with $N = 1024$, $\alpha = \sqrt{1-|\beta|^2}$, and $\beta = 0$, $0.02i$, and $0.04i$ corresponding to the solid black, dashed red, and dotted green curves, respectively.}
\end{center}
\end{figure}

Recently, \cite{Wong2015c} considered the effects of potential energy barriers hindering a quantum particle from searching for a marked vertex on the complete graph of $N$ vertices, as shown in Fig.~\ref{fig:complete}. To review, the vertices label computational basis states of an $N$-dimensional Hilbert space, and the $d = N-1$ directions from each vertex label a $d$-dimensional ``coin'' Hilbert space that is necessary to define a non-trivial quantum walk \cite{Meyer1996a,Meyer1996b}. Then the system $\ket{\psi}$ begins in an equal superpositions over both spaces:
\[ \ket{\psi_0} = \ket{s_v} \otimes \ket{s_c}, \]
where $\ket{s_v} = \sum_{i=1}^N \ket{i}/\sqrt{N}$ and $\ket{s_c} = \sum_{i=1}^d \ket{i}/\sqrt{d}$, and it evolves by repeated applications of
\begin{equation}
	\label{eq:U}
	U = (\alpha S + \beta I) (I_N \otimes C_0) (R_a \otimes I_d),
\end{equation}
where $\alpha$ and $\beta$ are the amplitudes of successfully and failing to tunnel through the potential barrier, $S$ is the flip-flop shift \cite{AKR2005} that causes the particle to hop from one vertex to another and then turn around, $C_0 = 2\ketbra{s_c}{s_c} - I_d$ is the Grover diffusion coin \cite{SKW2003}, and $R_a$ is an oracle that flips the sign of the amplitude at the marked vertex. The evolution of the success probability as $U$ is repeatedly applied is shown in Fig.~\ref{fig:prob_time_N1024}. Without the potential barrier (\textit{i.e.}, $\beta = 0$), the success probability reaches $1/2$ at $\pi\sqrt{N}/2\sqrt{2}$ applications of $U$ \cite{Wong2015b}. As $\beta$ increases, however, the max success probability decreases such that the $\Theta(\sqrt{N})$ runtime is retained when $\beta = O(1/\sqrt{N})$, and otherwise the algorithm performs no better than classical. Thus the hopping errors must not only decrease with $N$ for the search to be fast, but they must decrease sufficiently quickly.

In this paper, we show how to correct for these errors by modifying the phases that the coin $C_0$ and oracle $R_a$ use, and this recovers the quadratic speedup so long as $|\beta|$ does not approach $1$. To do this, we first reinterpret the search problem as an amplitude amplification algorithm. Then we choose the phases such that the amplification rotates to the marked vertex with high probability. Finally, we show that the $\Theta(\sqrt{N})$ runtime is restored.


\section{Amplitude Amplification}

\begin{figure}
\begin{center}
	\includegraphics{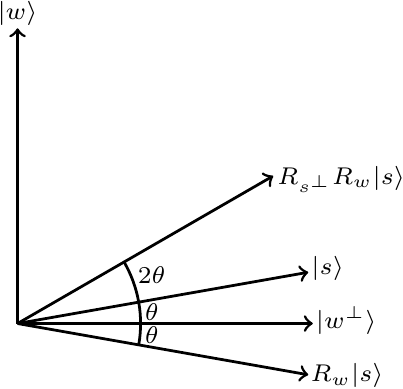}
	\vspace*{13pt}
	\fcaption{\label{fig:Grover_rotate} One application of Grover's iterate in the 2D subspace spanned by $\{ \ket{s}, \ket{w} \}$ results in a rotation of $2\theta$.}
\end{center}
\end{figure}

Figure~\ref{fig:complete} indicates that the system evolves in a 3D subspace spanned by $\{ \ket{ab}, \ket{ba}, \ket{bb}\}$, where $\ket{xy}$ indicates the equal superposition over the $x$ vertices pointing towards the equal superposition over the $y$ vertices \cite{Wong2015c}. Then in this basis, the operators in \eqref{eq:U} are
\begin{equation}
	S = \begin{pmatrix} 0 & 1 & 0 \\ 1 & 0 & 0 \\ 0 & 0 & 1 \end{pmatrix}, \enspace
	(I_N \otimes C_0) = \begin{pmatrix} 1 & 0 & 0 \\ 0 & -\frac{N-3}{N-1} & \frac{2\sqrt{N-2}}{N-1} \\ 0 & \frac{2\sqrt{N-2}}{N-1} & \frac{N-3}{N-1} \end{pmatrix}, \enspace
	(R_a \otimes I_d) = \begin{pmatrix} -1 & 0 & 0 \\ 0 & 1 & 0 \\ 0 & 0 & 1 \end{pmatrix}.
	\label{eq:operators}
\end{equation}
Using these, it is straightforward to show that
\[ \ket{\psi_{-1}} = \frac{1}{\sqrt{2N-3}} \begin{pmatrix}
	-\sqrt{N-2} \\
	-\sqrt{N-2} \\
	1 \\
\end{pmatrix} \]
is an eigenvector of \eqref{eq:U} since it is a $1$-eigenvector of $S$ and (trivially) $I$, and also $(-1)$-eigenvector of the product $(I_N \otimes C_0)(R_a \otimes I_d)$. Then consider the 2D subspace orthogonal to $| \psi_{-1} \rangle$, which is spanned by
\begin{gather*}
	\ket{s} = \frac{1}{\sqrt{N-1}} \ket{ba} + \sqrt{\frac{N-2}{N-1}} \ket{bb} = \begin{pmatrix} 0 \\ \frac{1}{\sqrt{N-1}} \\ \sqrt{\frac{N-2}{N-1}} \end{pmatrix}, \\
	\ket{w} = \frac{1}{\sqrt{2}} ( \ket{ab} - \ket{ba} ) = \frac{1}{\sqrt{2}} \begin{pmatrix} 1 \\ -1 \\ 0 \end{pmatrix}.
\end{gather*}
While these are both orthogonal to $| \psi_{-1} \rangle$, they are not orthogonal to each other, and we express the magnitude of their overlap as the sine of an angle $\theta$:
\[ \sin \theta = | \braket{s}{w} | = \frac{1}{\sqrt{2(N-1)}}. \]
Now let us find how the operators \eqref{eq:operators} act in this subspace. It is straightforward to show that
\[ S\ket{w} = -\ket{w}, \quad S | w^\perp \rangle = | w^\perp \rangle, \]
where $| w^\perp \rangle$ is the state orthogonal to $\ket{w}$, so $S$ is a reflection through $\ket{w}$. It is also straightforward to show that the coin and query together act by
\begin{align*}
	&(I_N \otimes C_0) (R_a \otimes I_d) \, \ket{s} = \ket{s}, \\
	&(I_N \otimes C_0) (R_a \otimes I_d) \, | s^\perp \rangle = -| s^\perp \rangle,
\end{align*}
so the coin and query together act as a reflection through $| s^\perp \rangle$. Thus the search operator $U$ \eqref{eq:U} without errors is the product of two reflections:
\[ \left. U \right|_{\alpha = 1, \beta = 0} = \underbrace{\phantom( S \phantom)}_{R_a} \underbrace{(I_N \otimes C_0) (R_a \otimes I_d)}_{R_{s^\perp}}. \]
This is simply Grover's iterate \cite{Grover1996, Aharonov1999}, which is illustrated in Fig.~\ref{fig:Grover_rotate}, with the reflections swapped, which does not affect the asymptotic search probability. Thus $\pi/4\theta \approx \pi\sqrt{N}/2\sqrt{2}$ applications of $U$ rotates $\ket{s}$ to $\ket{w}$ with probability near $1$. Since the initial equal superposition 
\[ \ket{\psi_0} = \frac{1}{\sqrt{N}} \left( \ket{ab} + \ket{ba} + \sqrt{N-2} \ket{bb} \right) \]
is approximately $\ket{s}$ for large $N$, the system roughly reaches a success probability $1/2$ (from the $\ket{ab}$ term in $\ket{w}$) in $\pi\sqrt{N}/2\sqrt{2}$ steps, in agreement with \cite{Wong2015b}.

Now with potential barriers, the shift/identity term multiplies by a different phase:
\begin{align*}
	&(\alpha S + \beta I) \ket{w} = -(\alpha - \beta) \ket{w} = -e^{-i\phi} \ket{w}, \\
	&(\alpha S + \beta I) | w^\perp \rangle = (\alpha + \beta) | w^\perp \rangle = e^{i\phi} | w^\perp \rangle, 
\end{align*}
where we parameterized $\alpha = \cos\phi$, $\beta = i\sin\phi$. Since the global phase does not matter, this is equivalent to acting by $-e^{-2i\phi}$ on $\ket{w}$ and doing nothing to $| w^\perp \rangle$. When $\phi = O(1/\sqrt{N})$, this deviation from an exact phase flip is insufficient alter the $\Theta(\sqrt{N})$ runtime, but when $\phi$ is scales larger, the algorithm is no better than classical \cite{Wong2015c}.


\section{Correcting Errors}

We now show how to compensate for effective $-e^{-2i\phi}$ phase that $(\alpha S + \beta I)$ applies to $\ket{w}$ by adjusting the coin and query operators, $C_0$ and $R_a$. In particular, we now use
\begin{gather*}
	C_0' = (1 + e^{i\eta}) \ketbra{s_c}{s_c} - I_d, \\
	R_a' \ket{a} = -e^{-i\eta} \ket{a}
\end{gather*}
for some phase $\eta$ that we must determine. Then the coin flip and oracle together are
\[ (I_N \otimes C_0')(R_a' \otimes I_d) = \begin{pmatrix}
	-1 & 0 & 0 \\
	0 & - \frac{N-2-e^{i\eta}}{N-1} & \frac{(1+e^{i\eta})\sqrt{N-2}}{N-1} \\
	0 & \frac{(1+e^{i\eta})\sqrt{N-2}}{N-1} & \frac{(N-2)e^{i\eta}-1}{N-1} \\
\end{pmatrix}. \]
Since $| \psi_{-1} \rangle$ is still a $(-1)$-eigenvector of this, we again consider the 2D subspace spanned by $\{\ket{s}, \ket{w}\}$. It is straightforward to show that
\begin{align*}
	&(I_N \otimes C_0') (R_a' \otimes I_d) \ket{s} = e^{i\eta} \ket{s}. \\
	&(I_N \otimes C_0') (R_a' \otimes I_d) | s^\perp \rangle = -| s^\perp \rangle.
\end{align*}
But since the global phase does not matter, this is equivalent to multiplying the phase of $| s^\perp \rangle$ by $-e^{-i\eta}$ and doing nothing to $\ket{s}$.

\begin{figure}
\begin{center}
	\includegraphics{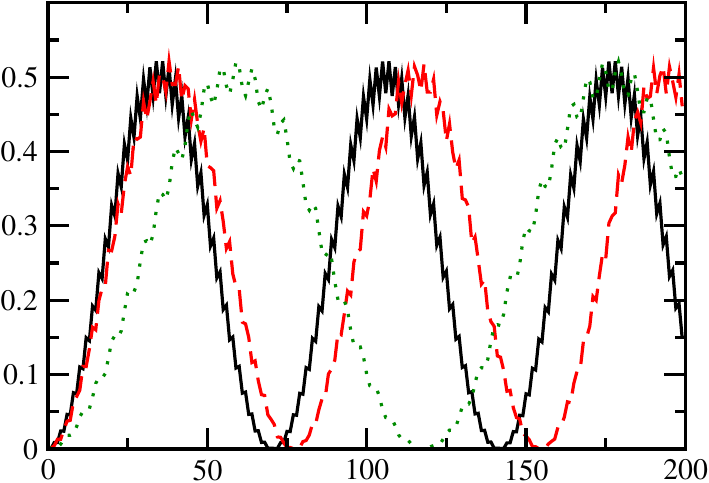}
	\vspace*{13pt}
	\fcaption{\label{fig:prob_time_N1024_corrected} Success probability as a function of the number of applications of $U'$ (search corrected for potential barriers) for search with $N = 1024$, $\alpha = \sqrt{1-|\beta|^2}$, and $\beta = 0$, $0.4i$, and $0.8i$ corresponding to the solid black, dashed red, and dotted green curves, respectively.}
\end{center}
\end{figure}

Thus the action of the modified search operator is
\[ U' = \underbrace{(\alpha S + \beta I)}_{-e^{-2i\phi} \text{ on } \ket{w}} \underbrace{(I_N \otimes C_0') (R_a' \otimes I_d)}_{-e^{-i\eta} \text{ on } \ket{s^\perp}}. \]
H\o{}yer derived a condition on these two phases if and only if amplitude amplification rotates the state $\ket{s}$ to $\ket{w}$ with certainty \cite{Hoyer2000}, which in our notation is
\[ \tan \left( \frac{-2\phi}{2} \right) = \tan \left( \frac{\eta}{2} \right) (1 - 2 \sin^2\theta). \]
Solving for $\eta$, which tells us what phase the adjusted coin $C_0'$ and query $R_a'$ should use, we get
\[ \eta = -2 \tan^{-1} \left[ \tan \left( \phi \right) \frac{N-1}{N-2} \right] \approx -2\phi. \]
With this choice of $\eta$, Fig.~\ref{fig:prob_time_N1024_corrected} shows the success probability as we repeatedly apply $U'$ with $N=1024$ and $\beta = 0$, $0.4i$, and $0.8i$. Even with these strong potential barriers, our correction still causes the system to rotate from $\ket{\psi_0} \approx \ket{s}$ to $\ket{w}$ with probability near $1$, as expected, which results in a success probability of $1/2$ from the $\ket{ab}$ term in $\ket{w}$.

As $\beta$ increases, however, the algorithm slows down. We can find the runtime by determining the angle of rotation $\sigma$ (which was $2\theta$ for the barrier-free case in Fig.~\ref{fig:Grover_rotate}) using H\o{}yer's result \cite{Hoyer2000}: $\sin\sigma = | \langle w | U' | \psi_0 \rangle |$, where $\ket{\psi_0} = \ket{s_v} \otimes \ket{s_c}$ is the initial equal superposition state. Evaluating this inner product with $\eta = -2\phi$,
\[ \langle w | U' | \psi_0 \rangle = \frac{(1+e^{-2i\phi})e^{-i\phi}}{\sqrt{2N}}. \]
The magnitude of this is
\[ \sin\sigma = | \langle w | U' | \psi_0 \rangle | = \sqrt{\frac{1+\cos2\phi}{N}}. \]
We want to rotate by a total amount roughly $\pi/2$, so if the number of applications of $U'$ is $t_*$, then we want $\sigma t_* = \pi/2$, or
\[ t_* = \frac{\pi}{2\sigma} = \frac{\pi}{2 \sin^{-1} \left(\sqrt{\frac{1+\cos2\phi}{N}}\right) } \approx \frac{\pi}{2 \sqrt{1+\cos2\phi}} \sqrt{N} \]
for large $N$. This runtime agrees with Fig.~\ref{fig:prob_time_N1024_corrected}, \textit{e.g.}, with $N = 1024$ and $\beta = 0.8i \Rightarrow \phi = \sin^{-1}(0.8)$, we get $t_* = 59$, which corresponds to the peak in success probability.

Using this runtime formula, we now show that the quadratic quantum speedup is recovered when $\phi$ does not approach $\pi/2$. First, when $\phi$ scales less than a constant, then $\cos2\phi \approx 1$ for large $N$, so the runtime is simply the barrier-free $\pi\sqrt{N}/2\sqrt{2}$. Furthermore, when $\phi$ scales as a constant (less than $\pi/2$ so the particle still hops), then while the runtime is slower, it still has the same $\Theta(\sqrt{N})$ scaling. Thus we have corrected for potential barriers by changing the phases of the coin flip $C_0$ and query $R_a$ in such a way that amplitude amplification recovers the full quadratic speedup, so long as the potential barriers get no worse as the problem grows (and assuming $\phi \ne \pi/2$ so that the particle still hops).

In the other extreme when $\phi$ approaches $\pi/2$, we have $\phi = \pi/2 - \delta$ with $\delta \to 0$. In this case, the particle mostly stays put, and one can consider it an ``error'' for the particle to tunnel to another vertex. Then our formula for $t_*$ yields a runtime of $\pi\sqrt{N} / 2\sqrt{2}\delta$. So when $\delta$ scales less than $1/\sqrt{N}$, the algorithm still has some improvement over classical. Thus rather than quantifying how small potential barriers must be to allow fast search, this quantifies how high potential barriers must be to stop the search. Finally, when $\delta = 0$ so that $\phi = \pi/2$, the particle does not hop at all, and $t_*$ is infinite, as expected, since the state is never rotated to $\ket{w}$.


\section{Continuous-Time Quantum Walk}

For completeness, we briefly state the correction for continuous-time quantum walks with potential barriers, which was given in \cite{Wong2015c}. If the potential barriers lower the transition amplitude of the particle by $\epsilon$, then the jumping rate $\gamma$ can be adjusted from its barrier-free value of $1/N$ to
\[ \gamma = \frac{1}{N(1-\epsilon)}, \]
and the walk searches with probability $1$ in time $\pi\sqrt{N}/2$ as if there were no potential barriers.


\section{Conclusion}

In the standard formulation of quantum walk search, even small potential barriers affecting the particle's hop can eliminate the quadratic quantum speedup. We have shown, however, that these errors can be corrected for by modifying the phases of the coin and query operators. By interpreting the quantum walk as an amplitude amplification algorithm, we find the precise phases to use and show that the quadratic speedup is restored. 

This approach to correcting errors from potential barriers should be applicable to a variety of quantum walk algorithms, so long as the evolution approximately occurs in a 2D subspace admitting rotations as in Grover's algorithm in Fig.~\ref{fig:Grover_rotate}, such as element distinctness \cite{Ambainis2004}. For algorithms where this does not hold, how to correct for potential barriers remains an open question, as does the effect of the potential barriers themselves.

Throughout this paper, we have assumed that the potential barriers are fixed and identical between every pair of vertices. Physically, however, they may be non-uniform or fluctuate randomly. The effects of these generalizations are open questions, as well as how to correct for them. They would also likely break the symmetry that makes this work tractable.


\nonumsection{Acknowledgements}
\noindent
This work was supported by the European Union Seventh Framework Programme (FP7/2007-2013) under the QALGO (Grant Agreement No.~600700) project, and the ERC Advanced Grant MQC.


\nonumsection{References}
\noindent
\bibliography{refs}
\bibliographystyle{qic}

\end{document}